\documentclass[prx,superscriptaddress,aps, twocolumn, reprint, showpacs, longbibliography, footnoography]{revtex4-2}
\usepackage{graphicx}
\usepackage{amssymb}   % for math
\usepackage{amsfonts}
\usepackage{amsmath}
\usepackage{dsfont}
\usepackage{natbib}
\usepackage{dcolumn}   % needed for some tables
\usepackage{bm}        % for math
\usepackage[mathscr]{eucal}
\usepackage[dvipsnames]{xcolor}
\usepackage[colorlinks, linkcolor=OrangeRed,citecolor=RoyalBlue,urlcolor=NavyBlue]{hyperref}
\usepackage[all]{hypcap} % let hyperlinks correctly point to figures rather than their captions;
\usepackage{mathtools}
\usepackage{dsfont}

\definecolor{myblue}{rgb}{0,0,0.75}

\newcommand{\ep}{\varepsilon}

\newcommand{\bra}[1]{\ensuremath{\left\langle#1\right|}}
\newcommand{\ket}[1]{\ensuremath{\left|#1\right\rangle}}

\newcommand\mean[1]{\ensuremath{\left\langle#1\right\rangle}}

\newcommand\lrp[1]{\left(#1\right)}
\newcommand\lrb[1]{\left[#1\right]}

\newcommand{\lra}{\;\; \Leftrightarrow \; \;}

\newcommand{\be}{\begin{equation}}
\newcommand{\ee}{\end{equation}}
\def\ba{\begin{aligned}}
\def\ea{\end{aligned}}
\newcommand{\bea}{\begin{eqnarray}}
\newcommand{\eea}{\end{eqnarray}}
\def\bes{\begin{subequations}}
\def\ees{\end{subequations}}
\def\bal{\begin{align}}
\def\eal{\end{align}}

\newcommand{\tr}{{\rm Tr}}

\usepackage{soul}
\usepackage{ulem} % needed for \sout-command

\usepackage{graphicx}
\usepackage{amsmath,amssymb,bm}
\usepackage{enumerate}
\usepackage{braket}        % Dirac notation

\graphicspath{{./figures/}}
\begin{document}
%\title{Non-Hermitian power-law random banded matrices: The good and the bad resonances}
\title{Stable many-body localization under random continuous measurements in the no-click limit}

\author{Giuseppe De Tomasi}
\affiliation{Department of Physics, University of Illinois at Urbana-Champaign, Urbana, Illinois 61801-3080, USA}
\author{Ivan M. Khaymovich}
\affiliation{Nordita, Stockholm University and KTH Royal Institute of Technology Hannes Alfv\'ens v\"ag 12, SE-106 91 Stockholm, Sweden}
\affiliation{Institute for Physics of Microstructures, Russian Academy of Sciences, 603950 Nizhny Novgorod, GSP-105, Russia}

\begin{abstract}
In this work, we investigate the localization properties of a paradigmatic model, coupled to a monitoring environment and possessing a many-body localized (MBL) phase. We focus on the post-selected no-click limit with quench random rates, i.e., random gains and losses. In this limit, the system is modeled by adding an imaginary random potential, rendering non-Hermiticity in the system. Numerically, we provide an evidence that the system is localized for any finite amount of disorder. To analytically understand our results, we extend the quantum random energy model (QREM) to the non-Hermitian scenario. The Hermitian QREM has been used previously as a benchmark model for MBL. The QREM exhibits a size-dependent MBL transition, where the critical value scales as $W_c\sim \sqrt{L} \ln{L}$ with system size and presenting many-body mobility edges. We reveal that the non-Hermitian QREM with random gain-loss offers a significantly stronger form of localization, evident in the nature of the many-body mobility edges and the value for the transition, which scales as $W_c\sim \ln^{1/2}{L}$ with the system size.
\end{abstract}
\maketitle
%%%%%%%%%%%%%%%%%%%%%%%%%%%%%
\section{Introduction}
%%%%%%%%%%%%%%%%%%%%%%%%%%%%%

Generic closed quantum many-body systems tend to thermalize under their own time evolution, erasing the microscopic details of their initial conditions~\cite{Deutsch1991, rigol2008thermalization, DAlessio2016ETH}. In such systems, the final steady state is characterized by a few set of globally conserved quantities and the system obey the law of statistical mechanics.

The investigation of non-equilibrium phenomena in quantum systems has unveiled new phases of matter. These phases usually lack counterparts in equilibrium states and could be crucial for enhancing the robustness of quantum computing.
To enhance coherence, several mechanisms have been suggested to impede thermalization, thus preserving critical quantum correlations essential for quantum computing. A prime example of this is the phenomenon known as many-body localization (MBL)~\cite{Basko06,gornyi2005interacting,Nandkishore_2015, Abanin_2019} which extends the Anderson localization to the many-body case. The strong quenched disorder prevents many-body systems from thermalizing, leading to localization. This ergodicity breaking manifests itself in the system's quantum dynamics~\cite{Canovi_2011, Richter_2022, Serbyn_2017, GOPALAKRISHNAN20201,Kotthoff_2021, Torres_2020, Bardarson_Sent_log(t), serbyn2013local,Serbyn_2014}. The local memory of the initial state is preserved during time evolution, attributed to a``robust'' form of an emergent integrability. Consequently, the system can be fully described by an extensive set of quasi-local integrals of motion~\cite{Imbrie2016, huse2014phenomenology, ROS_2015, Maksym_2013, serbyn2013local, De_To_2019, Serbyn_2014}. Later, alternative mechanisms for (weak) ergodicity breaking have been suggested, ranging from quantum many-body scars and Hilbert-space fragmentation~\cite{Serbyn2021, Moudgalya_scars_18, Strong_Frag_De_To_2019,Moudgalya_2022,Moudgalya2022, Moudgalya_Motrunich_2022} to the localization in lattice gauge theories~\cite{Smith_2017,Brenes_18, Russonmann_20, Karpov_21}.

Recently, there has been a growing interest in non-unitary quantum dynamics, both from theoretical and experimental perspectives. A notable instance is the recent exploration of monitored quantum systems~\cite{Chen_20,Tang_21,Li_18,Li_19}. In these systems, the unitary dynamics of typical quantum systems, which scramble and increase the entanglement of the state, are in competition with the continuous measurement of local observables, which locally projects the system into the low entanglement states.
This results in an intricate spreading of entanglement and correlations during the dynamics. Numerous studies have highlighted the existence of a distinct phase transition triggered by measurements~\cite{Skinner_2019,Zabalo_2022,Li_18,Li_19,Chan_19}, dubbed as the measurement-induced entanglement transition. This transition is commonly characterized by the scaling of entanglement dynamics and distinguishes between a volume-law phase, where entanglement scales with the volume of the system, and an area-law phase, where it grows as the boundary of the partition.

Static non-Hermitian quantum systems can describe some of the non-unitary dynamics mentioned above. For example, the non-Hermitian description arises naturally in the so-called ``no-click limit'', where the system is continuously monitored~\cite{Jacobs_2006, Paviglianiti_2023,Passarelli_2023,Le_Gal_2023, De_luca_2019, wiseman_milburn_2009, Ashida2020, Gopa_2021, Kazuki_2023}. Then one post-selects the quantum trajectories corresponding to no-measurement events. Several many-body systems, ranging from the transverse-field Ising model to the long-range Kitaev chain, show the existence of the entanglement transition, both in the dynamics and in the steady state properties~\cite{Le_Gal_2023, Turkeshi_2021, Turkeshi_2023, Zerba_2023, Paviglianiti_2023}.

Generically, the study of non-Hermitian many-body systems has emerged as a new paradigm for describing open, dissipative, and monitored systems. %with gain and loss. Generically,
The non-Hermitian systems uncover a rich phenomenology and the study of them is an active research front, describing unique effects, ranging from generalized topological phases and new forms of quantum criticality~\cite{RevModPhys_Flore_21}, to entanglement transitions that are not possible in the Hermitian counterparts. For instance, one of the most celebrated effects is a so-called non-Hermitian skin effect~\cite{Lin2023, Zhang_2022, Okuma_2023}, where eigenstates of a non-Hermitian system localize at the boundary of a lattice, arising due to %asymmetric or
non-reciprocal hopping across the system with open boundary condition. This effect challenges the conventional wisdom of the bulk-boundary correspondence in topological systems. Furthermore, it has been shown that such a skin effect suppresses entanglement propagation and thermalization and even induces an entanglement phase transition~\cite{Kawabata_2023}.

In general, non-Hermiticity is responsible for an increase in decoherence and might break  localization. Hatano and Nelson in their seminal work~\cite{Hatano_96} show that even in one dimension, where the short-range Hermitian systems are localized for any finite amount of disorder, strong non-reciprocity in hopping breaks the Anderson localization, creating  energy bands of delocalized states. The fate of the Hatano-Nelson model in the presence of interaction has been recently studied in several works~\cite{Ueda_2019,Zhai_2020, Suthar_2022, Wang_2023, Li_2023, Bhaseen_2023}, showing the existence of a stable MBL phase for both, weak disorder and non-Hermiticity.

In this work, we focus on a paradigmatic model, hosting the MBL transition under random continuous monitoring in the so-called no-click limit. In this limit, the model maps to a non-Hermitian Hamiltonian having complex random fields, which can be seen as random gain and loss terms. We inspect several probes to the extension and the entanglement properties of their eigenstates. We provide numerical evidence that the system is localized for any finite amount of disorder in the thermodynamic limit. To grasp an analytical understanding of this robust localization, we inspect a non-Hermitian version of the MBL-proxy, namely, the quantum random energy model (QREM). Previously, the Hermitian QREM is known to be a ``mean-field'' model to describe more realistic MBL systems. Generalizing the self-consistent theory of localization to the non-Hermitian QREM, we show that the non-Hermiticity results in parametrically stronger localization than in its Hermitian counterpart.

The rest of the work is organized as follows.
In Sec.~\ref{Sec:model} we describe the non-Hermitian modification of a paradigmatic MBL model, followed by the numerical results in Sec.~\ref{Sec:numerics}.
Section~\ref{Sec:Analitical} provides the analytical consideration of a non-Hermitian version, Sec.~\ref{Subsec:nH_QREM}, of a proxy QREM model both for finite, Sec.~\ref{Subsec:finite_e}, and zero, Sec.~\ref{Subsec:zero_e}, energy density.
We conclude our consideration in Sec.~\ref{Sec:concl}.

%%%%%%%%%%%%%%%%%%%%%%%%%%%%%
\section{ Model $\&$ Methods}\label{Sec:model}
%%%%%%%%%%%%%%%%%%%%%%%%%%%%%
We consider the random-field Heisenberg chain of  size $L$
\begin{equation}\label{eq:ham_MBL}
H_{0} = \sum_{x}^L S^+_x S^-_{x+1}+h.c.+\Delta S^z_xS^z_{x+1} + W \sum_x^L \mu_x S^z_x,
\end{equation}
subject to a random monitoring environment in the "no-click" limit. $-1\leq\mu_x \leq 1$ is the random field and we use periodic boundary conditions $\text{S}_{L+1} = \text{S}_{1}$. We set the interaction $\Delta$ to be $\Delta = 1$ unless stated otherwise. The model conserves the total magnetization $M= \sum_x S^z_x$ and, we consider the largest sector with zero magnetization ($M=0$).

Without monitoring, the model is known to exhibit an MBL transition, separating an ergodic phase from a localized one, around $W_{MBL} \approx 4-6$ in the middle of its spectrum~\cite{Pal_2010, Luitz15, Bera_2015, Devakul_2015, Sierant_2020, GDT_2017, Bera_2017, GDT_mutual_2017, Dogger_18}. For $\Delta=0$, the system is Anderson localized for any finite amount of disorder $W>0$~\cite{Mott_1961, Anderson1958}.

We consider the limit of post-selection, also called forced measurement phase or no-click limit~\cite{Jacobs_2006, Paviglianiti_2023,Passarelli_2023,Le_Gal_2023, De_luca_2019, wiseman_milburn_2009, Ashida2020, Gopa_2021}, in which the system is modeled by a non-Hermitian Hamiltonian
\begin{equation}
\label{eq:H_moni}
H = H_{0} + i\sum_x \gamma_x L^\dagger_x L_x,
\end{equation}
where $L_x$ are random Lindblad operator $L_x = (2 S^z_x+1)$. $\gamma_x$ are random rates $|\gamma_x|\leq W$~\footnote{In the standard setting $\gamma_x$ are considered to be non-negative, thus, having only losses. We consider the case of gains and losses, balanced on average, though it can be considered as a shift of the overall energy by $W$.}. Through the Jordan-Wigner transformation, the non-Hermitian Hamiltonian is equivalent to the disorder spin-less Hubbard model subject to random gain and loss terms. The monitored system in the non-interacting limit, $\Delta=0$ is also localized under the effect of random gain and loss both in the short-range Anderson~\cite{Huang_2020_ALT,Huang_2020_spectral} and long-range~\cite{DeTomasi2022nonHerm_RP,DeTomasi2023nonHerm_PLRBM} settings. In Ref.~\onlinecite{Ueda_2019} a similar model with staggered gain and loss, but non-random has been considered to find the existence of an MBL phase at strong disorder.

To understand the fate of the MBL transition under monitoring of the model in Eq.~\eqref{eq:H_moni}, we investigate the localization properties of the right-eigenvector belonging to the middle of the spectrum of $H$~\footnote{We focus on the right eigenvectors, as we do not observe any significant difference with the left eigenvector, belonging to the middle of the spectrum.},
\begin{equation}
H |E \rangle = E |E\rangle.
\end{equation}

With this aim, we consider the half-chain bipartite entanglement entropy,
\begin{equation}\label{eq:S(L/2)}
S(L/2) =- \tr [\rho_{L/2} \ln{\rho_{L/2}}],
\end{equation}
with the reduced density matrix $\rho_{L/2}$ of the state in a half chain.
The ergodic behavior of $S$ is known to be represented by the so-called Page value $S_{Page} = (L \ln 2 - 1)/2$~\cite{Page_1993_Sent}, with some finite deviations, claimed later~\cite{Huang_NPB2019_universal_EE,Sent2020_Haque,Huang_2021_universal_EE,Vidmar2023_Sent,sarkar2023entanglement}.
Here and further, we omit the argument $(L/2)$, focusing only on the half-chain partitions. In general, we indicate the disorder and eigenstates average with an overbar on the considered quantity, i.e., $\overline{S}$. As a further probe, we consider fluctuations over a few eigenstates and disorder configurations $\overline{\delta S^2} \equiv \overline{S^2} - \overline{S}^2$.

In an extended phase, we expect $\overline{S}$ to scale linearly with system size $L$, while in the localized, $\overline S$ should have an area-law scaling, $\overline S\sim O(L^0)$. Instead, the fluctuations decay exponentially with system size in the ergodic phase and do not scale with $L$ in the localized phase.  These two behaviors are interpolated by diverging fluctuations, $\overline{\delta S^2} \sim L^2$, around the transition ($W\approx W_{MBL}$). In the Hermitian case, it has been observed that the finite-size estimate of the critical point $W_{MBL}$ shifts to the larger values with increasing system sizes $L$. Several numerical works using finite-size techniques have shown that this shift is consistent with the existence of a genuine MBL transition at a finite value~\cite{Luitz15,Mace_Laflorencie2019_XXZ,De_Luca_2013,Tikhonov2018MBL_long-range, GDT_mutual_2017, ABANIN_2023}. However, recently, other works have questioned these results~\cite{Vidmar_2020_noMBL,Sels_2021_obstruct,Sirker_2019} proposing that only a finite-size transition is possible, whose critical value is proportional to the system size $W_{MBL}\sim L$.

We investigate the spread of eigenstates in the Hilbert space, quantified by the inverse participation ratio
\begin{equation}
IPR_2 = \sum_{\underline{\sigma}} |\langle \underline{\sigma} | E\rangle |^4,
\end{equation}
where $\{\ket{\underline{\sigma}}\}$
is the full set of the $\sigma^z-$basis states and $\ket{E}$ is the eigenstate of the above Hamiltonian with the eigenvalue $E$. From $IPR_2$, we extract its fractal dimension $D_2$
\begin{equation}
\label{eq:D_2}
IPR_2 \sim (\text{dim} \mathcal{H})^{-D_2} \ ,
\end{equation}
where $dim \mathcal{H} = \binom{L}{L/2}$ is the dimension of the Hilbert space in the zero-magnetization sector. For ergodic states, $D_2\rightarrow 1$ in the thermodynamic limit, while generic non-ergodic states will have fractal behavior, meaning $0<D_2<1$. For the Hermitian case, it has been shown that $D_2\approx 1$ in the thermal phase and $0<D_2<1$ in the MBL phase~\cite{Luitz15,Mace_Laflorencie2019_XXZ,Tikhonov2018MBL_long-range,QIsing_2021}. It has also been argued that $D_2$ exhibits a jump at the transition. The fluctuations of $D_2$ over eigenstates and disorder configurations, $\delta D_2^2 = \overline{D_2^2}-\overline{D_2}^2$, is a remarkable fingerprint of the transition. $\delta D_2^2$ should be exponentially suppressed with $L$ in the ergodic phase ($\delta D^2_2 \sim O(e^{-\alpha L})$), only algebraically in the MBL phase ($\delta D^2_2 \sim O(L^{-1})$), and $\delta D^2_2$ stays finite ($\delta D^2_2 \sim O(1)$) at the critical point, due to the above jump in $D_2$. Further, we focus on the fluctuations of $\ln IPR_2$, $\overline{\delta \ln IPR_2^2} = \delta D_2^2 L^2$.

%%%%%%%%%%%%%%%%%%%%%%%%%%%%%
%\section{Results}
%%%%%%%%%%%%%%%%%%%%%%%%%%%%%
\section{Numerical Results}\label{Sec:numerics}
%%%%%%%%%%%%%%%%%%%%%%%%%%%%%
We start our investigation with the examination of the entanglement entropy and its fluctuations as a function of disorder strength $W$, focusing on the right eigenvectors from the middle of the spectrum of $H$ in  Eqs.~\eqref{eq:ham_MBL},~\eqref{eq:H_moni}.

\begin{figure}[t!]
\label{fig:XXZ_EE}
    \includegraphics[width=1.\columnwidth]{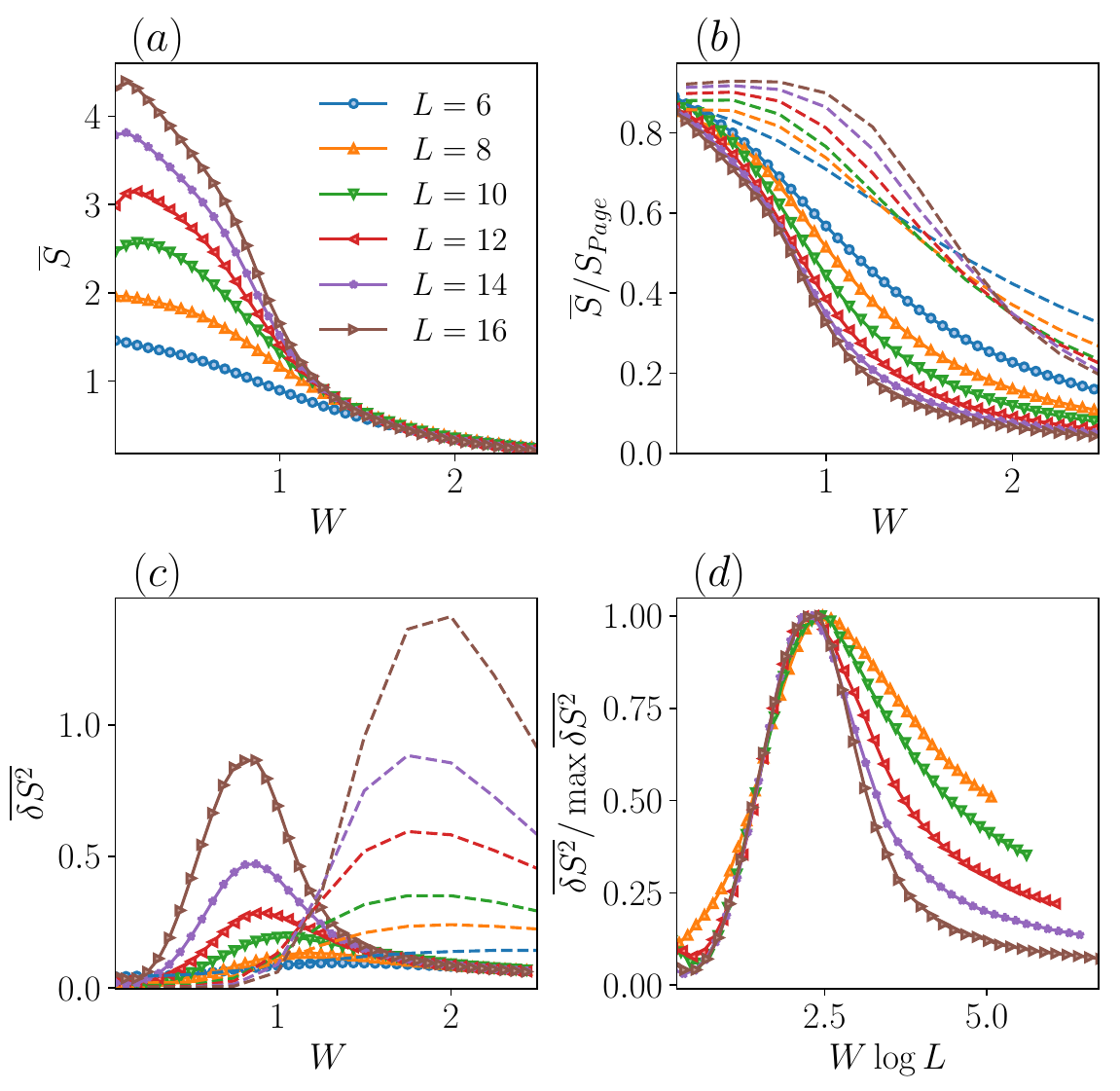}
    \caption{\textbf{MBL model: Entanglement entropy and its fluctuations in a non-Hermitian setting of Eq.~\eqref{eq:H_moni}:} (a)~$\overline{S}$ and (b)~$\overline{S}/S_{Page}$ as functions of $W$ for several system sizes $L$. (c)~bare entanglement fluctuations $\overline{\delta S^2}$ and (d)~its value, normalized by the maximum value.
    In panel~(d) the disorder strength has been renormalized by multiplying by $\ln{L}$. Dashed lines in panels (b), (c) show the corresponding measures for the equivalent Hermitian Hamiltonian, Eq.~\eqref{eq:H_Hermitian}.
    }
\end{figure}

Figures~\ref{fig:XXZ_EE}(a)-(b) depict the average half-partition entanglement entropy, Eq.~\eqref{eq:S(L/2)}, and the ratio $\overline{S}/S_{Page}$ (represented by solid lines), respectively. %Here, $S_{Page}$ refers to the entanglement entropy of a random state, defined as $S_{Page}= L/2\ln{2}-1/2$.
Here, the Page value $S_{Page}$ will be used as a reference value, being the expected value for ergodic states at infinite temperature (middle of the spectrum). The rest of the figure, Figs.~\ref{fig:XXZ_EE}(c)-(d) showcase the fluctuations of $S$.

Notably, $\overline{S}/S_{Page}$ decreases as the system size increases, and the peak of its fluctuations also appears to systematically move to a smaller value of $W$, see Fig.~\ref{fig:XXZ_EE}~\footnote{Note that in the case of non-reciprocal hopping, the peak of the entanglement fluctuations does not develop, see, e.g.,~\cite{Ueda_2019}}. This behavior may suggest the presence of a finite-size crossover between a volume-law phase and an area-law phase. However, the volume-law phase seems to vanish in the thermodynamic limit.

To support these observations, it is valuable to compare them with the Hermitian case, which is believed to exhibit the MBL transition. For a fair comparison between the Hermitian and the non-Hermitian settings described in Eq.~\eqref{eq:H_moni}, we examine the Hamiltonian, where the complex disorder is replaced by the real one
\begin{equation}
\label{eq:H_Hermitian}
H_{h} = H_{0} + \sum_x L^\dagger_x L_x,
\end{equation}
as depicted by dashed lines in Figs.~\ref{fig:XXZ_EE}(b) and~(c). First, we observe that in the Hermitian case the finite-size crossover between the thermal and the ergodic phases happens at larger disorder, indicating that the non-Hermitian part tends to localize the system. This is particularly evident in Figs.~\ref{fig:XXZ_EE}(b)%\TODO{No dashed lines in (a)!!}
, where the dashed lines present a crossing point in the proximity of the putative MBL transition. Indeed, $\overline{S}/S_{Page} \rightarrow 1$ as the system size grows for the weak disorder, while $\overline{S}/S_{Page} \rightarrow 0$ at larger disorder. %, please refer to the dashed lines in Fig.~\ref{fig:XXZ_EE}(b).
However, for the non-Hermitian case, we have $\overline{S}/S_{Page} \rightarrow 0$ with growing $L$ for all available $W$. The enhancement of localization due to non-Hermiticity is also visible in the fluctuation of $S$, cf. dashed and solid lines in Fig.~\ref{fig:XXZ_EE}(c). First, the maximum of $\overline{\delta S^2}$ is higher in the Hermitian case, second, it happens at stronger disorder, and finally, we observe the typical shifting of the maximum to the larger values of disorder only in the Hermitian case, but not in its non-Hermitian counterpart. We can conclude that this non-Hermiticity enhanced its localization properties.

Realizing that the system tends to the localization with increasing $L$, we analyze a finite-size scaling of $\overline{\delta S^2}$ as a function of $W$. As shown in Fig.~\ref{fig:XXZ_EE}(d), we have found a remarkably good collapse of the left wing of the $\overline{\delta S^2}$-maximum by rescaling the disorder strength, $W\rightarrow W\ln{L}$. This collapse provides an evidence that the parameter space hosting an extended phase shrinks with systems size as $\sim 1/\ln{L}$.
\begin{figure}[t!]
\label{fig:XXZ_I}
    \includegraphics[width=1.\columnwidth]{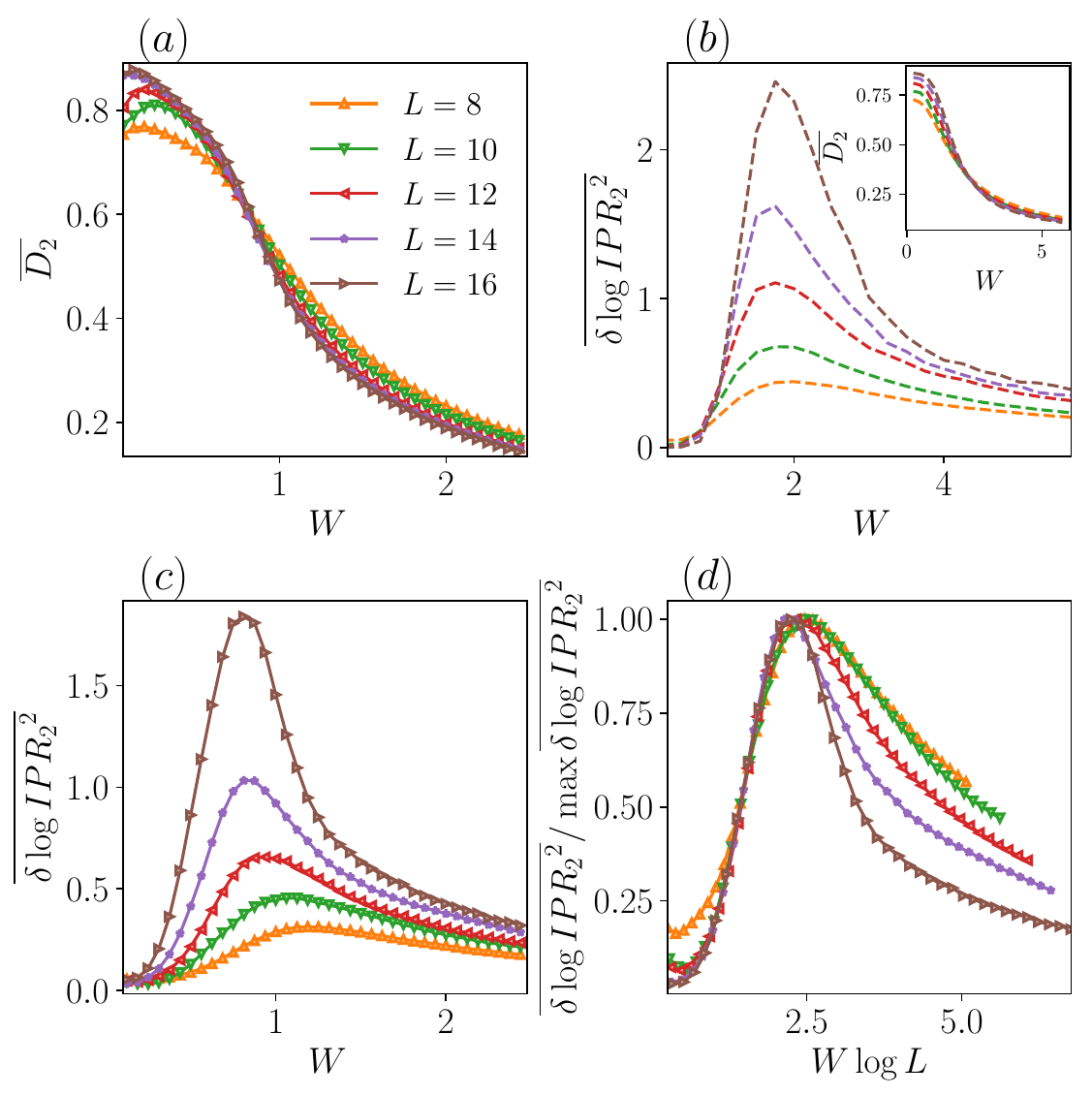}
    \caption{\textbf{MBL model: Fractal dimension $D_2$ and its fluctuations in a non-Hermitian setting of Eq.~\eqref{eq:H_moni}:} (a)~$\overline{D}_2$~as a function of $W$ for several $L$. (b)~shows the fluctuations of $\ln{IPR_2}$ in the Hermitian case of Eq.~\eqref{eq:H_Hermitian}, (inset) shows the corresponding $\overline{D}_2$ vs $W$.
    (c)~bare fluctuations $-\overline{\delta \ln{IPR_2}^2}$ of $-\ln{IPR_2}$ and (d)~its value, normalized by its maximum value. In panel (d) the disorder has been renormalized as follows $W\rightarrow W\ln{L}$.}
\end{figure}
%To further support our observations, we investigate the spread of the wave function through the fractal dimension $D_2$ in Eq.~\eqref{eq:D_2}. In Fig.~\ref{fig:XXZ_I}(a) and (c), we present the results for $\overline{D_2}$ and the fluctuations $\overline{\delta \ln{IPR_2}^2}$ as function of $W$ for several $L$. In agreement with the results of the entanglement entropy, the fractal dimension $D_2$ is strictly smaller than its ergodic value, i.e. $D_2\approx 1$, and the data are weakly dependent on $L$, except for quite small values of $W$. Indeed, $\overline{\delta \ln{IPR_2}^2}$ in Fig.~\ref{fig:XXZ_I}(c) has divergent value with $L$, approximately at the same value of $W$ in $\overline{\delta S^2}$. As in $\overline{\delta S^2}$, the maximum of $\overline{\delta \ln{IPR_2}^2}$ shifts to smaller values of $W$ with increasing $L$. Furthermore, while $\overline{\delta \ln{IPR_2}^2}\sim L$ for the strong disorder, as it should be in a localized phase, we do not see an exponential suppression of it at weak disorder, providing further an evidence of the non-existence of an extended phase in the thermodynamic limit.

To bolster our findings, we delve into the wave-function Hilbert-space spreading via the fractal dimension $D_2$, defined by Eq.~\eqref{eq:D_2} as depicted in Fig.~\ref{fig:XXZ_I}. Indeed, in Fig.~\ref{fig:XXZ_I}(a) and (c), we display the results for $\overline{D_2}$ and the fluctuations $\overline{\delta \ln{IPR_2}^2}$ as a function of $W$ for various values of $L$. Consistently with the above entanglement entropy data, the fractal dimension $D_2$ is significantly below its ergodic value, $D_2=1$, and the data shows only a mild dependence on $L$, and mostly for relatively small values of $W$. The IPR fluctuations, $\overline{\delta \ln{IPR_2}^2}$ in Fig.~\ref{fig:XXZ_I}(c) diverge with increasing $L$, approximately at the same $W$ value as $\overline{\delta S^2}$ in Fig.~\ref{fig:XXZ_EE}. Analogously to $\overline{\delta S^2}$, the peak of $\overline{\delta \ln{IPR_2}^2}$ flows to lower $W$ values, as $L$ increases. Moreover, while $\overline{\delta \ln{IPR_2}^2}\sim L$ is evident in the presence of strong disorder, as expected in a localized phase, there's no observed exponential suppression at weak disorder values. This further justifies the probable absence of an extended phase in the thermodynamic limit.

As we previously did by analyzing the entanglement entropy in Fig.~\ref{fig:XXZ_EE}, {it is crucial to compare this result with the results of the Hermitian case %it's pivotal to juxtapose this outcome with those of the Hermitian case
 in Eq.~\eqref{eq:H_Hermitian}. The corresponding data is presented in Fig.~\ref{fig:XXZ_I}(b), illustrating $\overline{\delta \ln{IPR_2}^2}$, while the inset
shows $\overline{D}_2$ vs $W$.
Supporting the insights from the entanglement entropy, the IPR data underscores the propensity of non-Hermitian components to enhance the system's localization. Conclusively, a robust convergence for the fluctuations of the fractal dimension is evident in Fig.~\ref{fig:XXZ_I}(d), attesting that the extended phase diminishes in the thermodynamic limit, scaling as $1/\ln{L}$.

In summary, through the examination of multiple complementary indicators, we present numerical proof that our monitored random-field Heisenberg model exhibits the localization in the thermodynamic limit, irrespectively to the disorder magnitude.

%%%%%%%%%%%%%%%%%%%%%%%%%%%%%%%%%%%%%%
\section{Analytical consideration}\label{Sec:Analitical}
%%%%%%%%%%%%%%%%%%%%%%%%%%%%%%%%%%%%%%
The aim of this section is to provide some analytical considerations that will help us to understand the enhancement of localization that we have observed numerically.
%%%%%%%%%%%%%%%%%%%%%%%%%%%%%
\subsection{Non-Hermitian QREM}\label{Subsec:nH_QREM}
%%%%%%%%%%%%%%%%%%%%%%%%%%%%%
Following the analysis presented in Ref.~\onlinecite{Laumann2014QREM}, we consider the QREM as a toy model that exhibits an MBL transition. The Hermitian QREM is defined as:
\begin{equation}
\label{eq:QREM}
    H_{hQREM} = \sum_i^L \sigma_i^x + W \sum_{\underline{\sigma}}^{2^L} E_{\underline{\sigma}}^R \ket{\underline{\sigma}}\bra{ \underline{\sigma}},
\end{equation}
where $\sigma_i^x$ is the $x-$Pauli matrix at site $i$, $\{ |\underline{\sigma} \rangle \}$ are the $2^L$ product states in the $\sigma^z-$basis and $E_{\underline{\sigma}}^R$ are independent identically distributed Gaussian random variables $\mathcal{N}(0, \sqrt{L})$.  From a Fock space point of view, the first term in Eq.~\eqref{eq:QREM} introduces jumps between $z$-spins configurations, which differ by one spin flip. Instead, the second term is composed of independently distributed random energies whose width scales as $\sqrt{L}$ to mimic many-body density of states, see Eq.~\eqref{eq:P(E)} below, given by the diagonal disorder and interaction terms. We can decompose the second term in Eq.~\eqref{eq:QREM}  in a string composed by $\sigma^z$ Pauli matrices,
\begin{gather}\label{eq:diag_sum_QREM}
\sum_{\underline{\sigma}}^{2^L} E_{\underline{\sigma}}^R |\underline{\sigma} \rangle \langle \underline{\sigma}| = \sum_n^L \sum_{i_1,\cdots , i_n} J_{i_1\cdots i_n} \sigma^z_{i_1}\cdots \sigma_{i_n}^z \ .
\end{gather}
After this decomposition, we can interpret the QREM as a quantum Ising type chain subject to all possible $n$-body $\sigma^z$ interactions, with $n$ from $1$ to $L$.
The main difference of the QREM model with the Quantum Ising model~\cite{Imbrie2016} is that in the latter sum, Eq.~\eqref{eq:diag_sum_QREM}, only the terms with $n=1$ and $2$ survive, and coefficients $J_i, J_{i,i+1} \sim O(1)$, leading to the strong correlations between the many-body diagonal energies $E_{\underline{\sigma}}^R$.

The QREM phase diagram of has been investigated in Ref.~\onlinecite{Laumann2014QREM}, finding the existence of a finite-size metal-insular transition of mid-spectrum states at $E=0$, whose critical value scales as $W_c \sim \sqrt{L} \ln{L}$. Furthermore, $H_{hQREM}$ has stable energy mobility edges at finite $W$.

In this section, we inspect the fate of the QREM under the presence of random non-Hermitian terms
\begin{equation}\label{eq:nH_QREM}
        H_{nQREM} = \sum_i^L \sigma_i^x + W \sum_{\underline{\sigma}}^{2^L} E_{\underline{\sigma}}^R |\underline{\sigma} \rangle \langle \underline{\sigma}| + i W' \sum_{\underline{\sigma}}^{2^L} E_{\underline{\sigma}}^I |\underline{\sigma} \rangle \langle \underline{\sigma}| ,
\end{equation}
where $E_{\underline{\sigma}}^I$ are also are independent identically distributed Gaussian random variables %\TODO{Do we need this 'a' here?}
$\mathcal{N}(0, \sqrt{L})$.

Now, we are studying the phase diagram of the non-Hermitian QREM. In the following discussion, we will demonstrate how the localization transition in the QREM, at both finite and zero energy density, is influenced by non-Hermitian gain-loss disorder.

For clarity, let's first revisit the Hermitian case.

In the Hermitian case, $E_\sigma^I \equiv 0$ and real part $E_\sigma^R$, is normally distributed. In particular, for the finite density energy $\epsilon = E/L$, we have for its probability
\be\label{eq:P(E)}
P(E_\sigma^R/L = \ep)d \ep = e^{-L \ep^2}\sqrt{\frac{L}{\pi}}d\ep \ ,
\ee
which gives the typical energy scale, given by
\be\label{eq:E_typ}
E_{typ}\sim L^{1/2} \ .
\ee

On the localized side of the transition at large enough $W$, to the leading order in $1/W$ the wave-function amplitude at a certain Hilbert/Fock space node $b_n$ at the Hamming distance $n$ from the wave-function maximum $\psi(0)$ is
\be\label{eq:FSA_psi}
\psi(b_n) \simeq W^{-n}\sum_{p\in \Pi_n} \prod_{i\in p} \frac{1}{E_0 - E_i} \ ,
\ee
where $\Pi_n$ is the set of all $n!$ shortest directed paths between two Hilbert space nodes at the distance $n$, $p$ runs over all such paths, and $i$ runs over all nodes on the path $p$.

The main steps to find the phase diagram of $H_{hQREM}$ as the  following~\cite{Laumann2014QREM}:
\begin{itemize}
    \item First, we find the probability $p$ to have a resonance at a node $b_n$ summed over all paths at the finite-energy density, $E_0/L = \ep>0$, or calculated for a certain path $p$ for the zero energy density,  $\ep\to 0$;
    \item Next, we calculate the probability $P_n = (1-p)^{\mathcal{N}}$ to have no resonances in all $\mathcal{N} = \lrp{L\atop n}$ nodes $b_n$ for $\ep>0$ (all $\mathcal{N} = \lrp{L\atop n} n!$ paths between $0$ and $b_n$ for $\ep=0$),
    \item Then, the saddle-point approximation for $P_n \sim e^{-e^{M f(\sigma W)}}$ is calculated over the large parameter $M=L$ ($M=n$) with the renormalized $W$ via $\sigma = \ep$ ($\sigma = \sqrt{\pi L}/2$) for $\ep>0$ ($\ep=0$).
    \item Finally, the localization transition or the mobility edge is associated with the appearance of the first resonance when $f$ obtains the first positive value.
\end{itemize}

Here we repeat the same argumentation for the non-Hermitian QREM model with gain-loss complex disorder potential.

%%%%%%%%%%%%%%%%%%%%%%%%%%%%%%%%%%%%%%
\subsection{Finite energy density, $E_0/L = \ep >0$}\label{Subsec:finite_e}
%%%%%%%%%%%%%%%%%%%%%%%%%%%%%%%%%%%%%%
In {the case of finite energy density, the typical energy at first $n-1$ nodes of the path before the resonance is of the order of~\eqref{eq:E_typ} and can be neglected in~\eqref{eq:FSA_psi} with respect to $E_0\sim L$. Note that for this case the difference between QREM and the quantum Ising model, given by the correlations in~\eqref{eq:diag_sum_QREM} is not important, as the entire diagonal term can be neglected with respect to the eigenenergy $E_0$. Thus, the results, derived below for QREM should be the same for the many-body problem, like the quantum Ising model.

The above approximation gives the contribution from each path to each of $n$ neighbors $a_{n-1}$ of $b_n$ at the distance $n-1$ from the wave-function maximum
\be
\psi(a_{n-1},p) \simeq \lrp{\frac{1}{W \ep L}}^{-(n-1)}
\ee
which have the same sign and, thus, can be summed coherently and the same for all $a_{n-1}$
\be
\psi(a_{n-1}) \simeq \sum_{p\in \Pi_n} \psi(b_n,p) \simeq n! \lrp{\frac{1}{W \ep L}}^{-(n-1)} \equiv \psi_{n-1} \ .
\ee

For resonance at $b_n$,  one must assert that the resulting contribution to
\be
\psi(b_n) \simeq \frac{n}{W \delta_n} \psi_{n-1}
\ee
should be large $|\psi(b_n)|>1$. It can happen only if the energy difference $\delta_n = |E_0 - E_{b_n}|$ is small enough
\be
\delta_n < \delta_c \equiv\frac{n}{W} \psi_{n-1} \simeq 2 L \ep \sqrt{2\pi n} \lrp{\frac{n}{e \ep W L}}^{n} \ .
\ee
%In the latter, we have used the Striling's formula for $n!$.
Note that for all positive integer $n$ the r.h.s. of the latter is small compared to the width~\eqref{eq:E_typ} of the distribution~\eqref{eq:P(E)}.

Thus, the probability $p$ to have such a resonance is given by the following expression
\be\label{eq:p}
p = \int_{\delta_n < \delta_c} P(E_n) d E_n \ .
\ee
For the Hermitian case of~\cite{Laumann2014QREM} it immediately gives
\be
p_H = P(\ep) \delta_c \simeq 2\sqrt{2 x} \ep L \cdot e^{-L \ep^2 - x L \ln\lrb{e W \ep/x}} \ ,
\ee
where $x = n/L$. Up to polynomial corrections in $L$, we will focus on the leading saddle-point approximation and neglect the prefactor.

In the non-Hermitian case, the integral in Eq.~\eqref{eq:p} is two-dimensional over independent $E_n^R$ and $E_n^I$, distributed according to~\eqref{eq:P(E)}, thus, within the same approximation $\delta_n\ll \sqrt{L}$ one obtains
\be
p_{nH} = p_H^2 \ .
\ee

The probability $P_n$ to not have resonances in all $\mathcal{N} = \lrp{L\atop n}$ directed paths connecting the stating point with $b_n$ is given by product
\be\label{eq:P_n}
P_n = (1-p)^{\mathcal{N}} \simeq e^{-p \mathcal{N}} \ ,
\ee
where
\be
p \mathcal{N} \simeq e^{L f(x,p)} \ .
\ee
The function $f(x,p)$ differs for Hermitian and non-Hermitian cases,
\bea
f_H &=& -x\ln x - (1-x)\ln (1-x) - \ep^2 - x\ln\lrb{e W \ep/x} \nonumber \\ &=&  - (1-x)\ln (1-x) - \ep^2 - x\ln\lrb{e W \ep} \ , \\
f_{nH} &=& -x\ln x - (1-x)\ln (1-x) - 2\ep^2 - 2x\ln\lrb{e W \ep/x} \nonumber \\ &=&  x\ln x - (1-x)\ln (1-x) - \ep^2 - 2x\ln\lrb{e W \ep} \ .
\eea
At small $\ep W \ll 1$ the maxima of the above functions over $x$ appear at
\be
x^*_H = 1 - \frac{1}{\ep W} \ , \quad
x^*_{nH} = 1 - \lrp{\frac{1}{\ep W}}^2 \ ,
\ee
leading to the solutions of the equation $f(x^*,p) = 0$ given by
\be
W_H^{-1} = \ep + \sqrt{2}\ep^2 + O\lrp{\ep^3}  \ , \quad
W_{nH}^{-1} = \ep e^{1+\ep^2} \ .
\ee
This is the main result for the many-body mobility edge, showing that at the same energy density, the critical disorder in the non-Hermitian case is suppressed by a factor $e^{1+\ep^2}$.

The corresponding distance $n^* = x^* L$ where the first resonance appear is given by
\be
x^*_H = \sqrt{2} \ep + O(\ep^2) \ , \quad
x^*_{nH} = 1 - e^{-2(1+\ep^2)} \ .
\ee
Here, the main difference between the Hermitian and non-Hermitian cases is that the first resonance at zero energy density in the former case appears at very short distances, where $n^*/L\to 0$, approaches zero, while in the latter case, it remains finite, resulting in sharper mobility edges.

%%%%%%%%%%%%%%%%%%%%%%%%%%%%%%%%%%%%%%
\subsection{Zero energy density, $E_0/L = \ep =0$}\label{Subsec:zero_e}
%%%%%%%%%%%%%%%%%%%%%%%%%%%%%%%%%%%%%%
Unlike the previous case, here the terms of the different paths to $b_{n}$ are not coherent due to the random signs of the denominators in Eq.~\eqref{eq:FSA_psi} and this forces us to consider all $\mathcal{N} =\lrp{L\atop n} n!$ paths independently and assume that the first resonance appears only on one of them~\footnote{Note that it is not the case for the MBL problems, where the resonances appear simultaneously due to the strong correlations in the diagonal energy, Eq.~\eqref{eq:diag_sum_QREM}.}.

Following~\cite{Laumann2014QREM}, we introduce a random variable
\be
y_i = \ln\lrp{\sigma/|E_i|} \lra |E_i| =  \sigma e^{-y_i}\ ,
\ee
where $\sigma = \sqrt{\pi L}/2$ is chosen in such a way to make the distribution $P(y_i\gtrsim 1)$ to be parameter free.

For the Hermitian case
\be
P(y_i) = P(\ep_i) \frac{d\ep_i}{d y} = e^{-y_i - \frac{\pi}{4} e^{-2 y_i}} \lra P(y_i\gtrsim 1) \simeq e^{-y_i} \ ,
\ee
while in the non-Hermitian case, the results are more involved
\begin{multline}
P(y_i) = \int \frac{e^{-(E_R^2+E_I^2)/L}}{\pi L} \delta\lrp{y_i +\frac12 \ln\lrb{\frac{E_R^2+E_I^2}{\sigma^2}}} dE_R dE_I =\\=
\frac{e^{-\frac{\pi}{4} e^{-2 y_i}}}{\pi L}\int_{-\sigma e^{-y_i}}^{\sigma e^{-y_i}} \frac{\partial E_R(y_i,E_I)}{\partial y_i} d E_I  =\\=
\frac{2 e^{-\frac{\pi}{4} e^{-2 y_i}}}{\pi L} \sigma^2 e^{-2y_i}\int_{-1}^1 \frac{dz}{\sqrt{1-z^2}} = \\ =
\pi e^{-\frac{\pi}{4} e^{-2 y_i}-2y_i}
\end{multline}
where we used the expression $E_R(y_i, E_I) = \sqrt{\sigma^2 e^{-2y_i} - E_I^2}$ and $z = E_I e^{y_i}/\sigma$.

The first resonance appears at $b_n$ as soon as
\be\label{eq:Y_resonance}
\ln |\psi(b_n)| > 1 \lra
Y_n = \sum_{i=1}^n y_i > n \ln \lrp{\sigma W} \equiv Y_c \ .
\ee

In our approximation of $|y_i|\gtrsim 1$, the distribution of $P(Y)$ reads differently for Hermitian
\be
P_H(Y) \simeq \frac{Y^{n-1}}{(n-1)!} e^{-Y} \simeq \lrp{\frac{Y e}{n}}^{n} e^{-Y}
\ee
and non-Hermitian cases:
\be
P_{nH}(Y) \simeq \frac{\lrp{2 Y}^{n-1}}{2(n-1)!} e^{-2Y} \simeq \lrp{\frac{2Y e}{n}}^{n} e^{-2Y} \ .
\ee

For large enough $Y_c = n \ln\lrp{\sigma W}\gg 1$ the probability to have a resonance~\eqref{eq:Y_resonance} is given by
\be
p = P(Y_c) \cdot\lrp{1+ O(Y_c^{-1})} \ ,
\ee
thus, the probability to have no resonances for all $\mathcal{N}=\lrp{L\atop n} n!\simeq L^n$ paths reads exactly as~\eqref{eq:P_n}, leading to the following saddle-point approximation over $n$
\be
p \mathcal{N} = e^{n f(\sigma W)} \ ,
\ee
with $f$ being different for two above cases via the constant $c$
\begin{multline}
f(\sigma W) \equiv \ln L +\frac1n \ln\lrb{P(Y_c)}=\\=
\ln(e L) - \ln\lrp{\sigma W}^c + \ln\lrb{\ln\lrp{\sigma W}^c} \ .
\end{multline}
$c=1$ for Hermitian and $c=2$ for non-Hermitian cases.

\begin{figure}[t!]
\label{fig:QREM_EE}
    \includegraphics[width=1.\columnwidth]{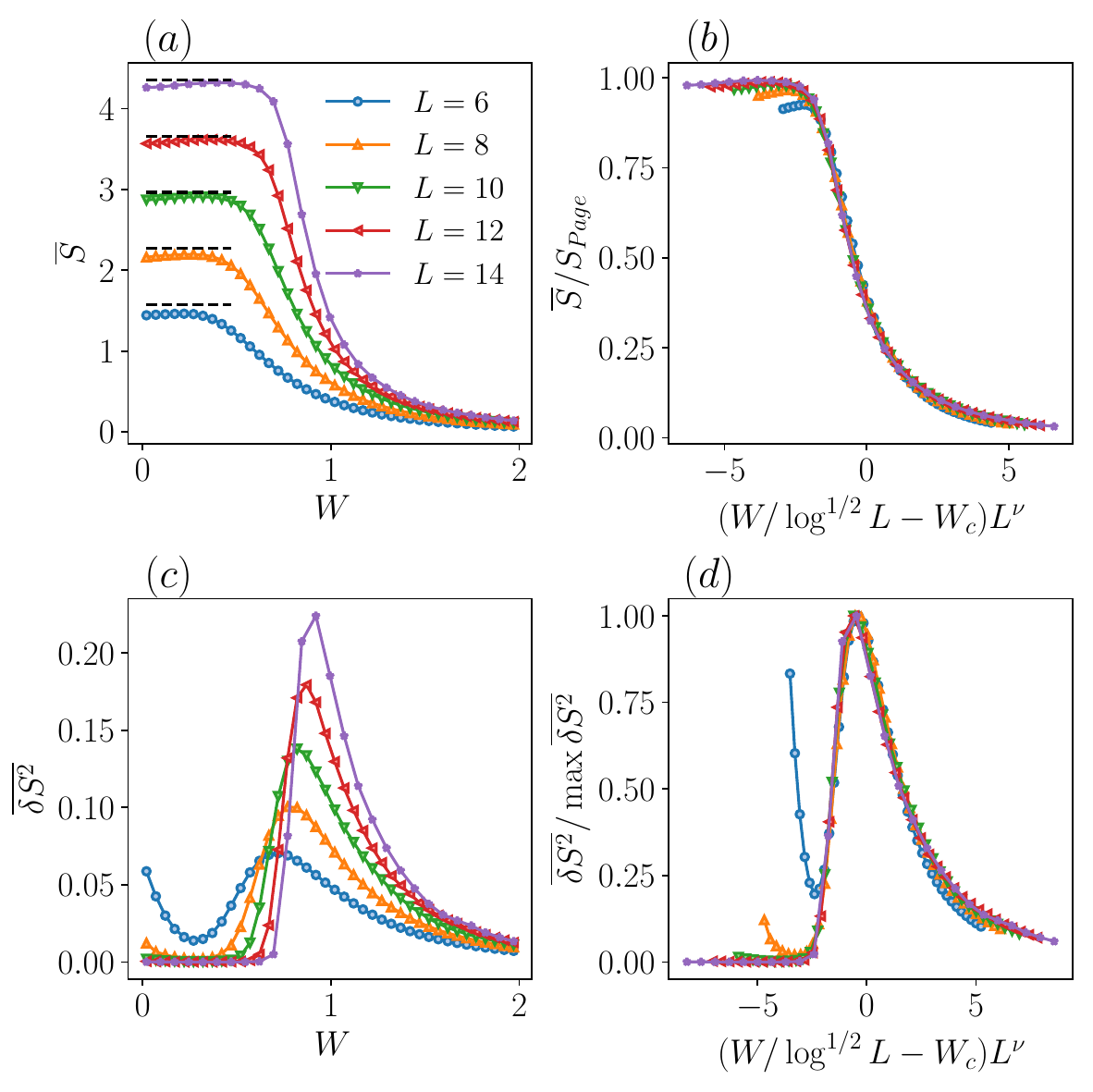}
    \caption{\textbf{QREM model: Entanglement entropy and its fluctuations in a non-Hermitian setting of Eq.~\eqref{eq:nH_QREM}:} (a)~$\overline{S}$, as a function of $W$ for several $L$; (b)~finite-size collapse of $\overline{S}/S_{Page}$. The panel shows that the critical value shifts to larger values as $\ln^{1/2}(L)$. (c)~Fluctuations $\overline{\delta S^2}$ of $S$ and (d)~finite-size collapse of $\overline{\delta S^2}$, normalized by its maximal value. In both panels (b) and (d) the finite-size collapse gives $W_c\approx 0.5$ and $\nu\approx 1$.
    }
\end{figure}
\begin{figure}[b!]
\label{fig:QREM_I}
    \includegraphics[width=1.\columnwidth]{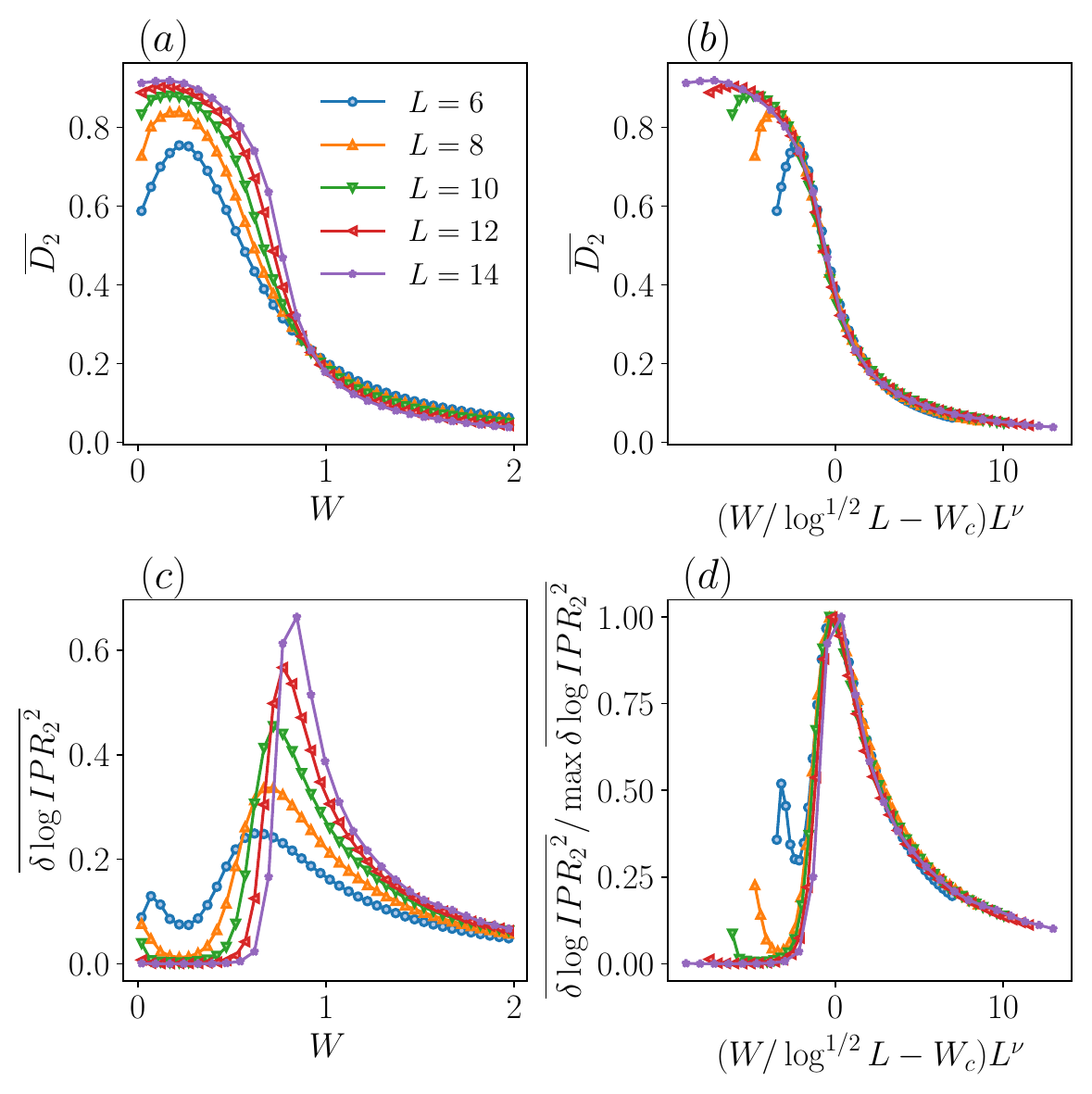}
    \caption{\textbf{QREM model: Fractal dimension $D_2$ and its fluctuations in a non-Hermitian setting of Eq.~\eqref{eq:nH_QREM}:} (a)~$\overline{D}_2$ as a function of $W$ for several $L$; (c)~fluctuations $\overline{\delta \ln{IPR_2}^2}$ of $-\ln{IPR_2}$. (b),(d) Finite-size collapses of $\overline{D}_2$ and $\overline{\delta \ln{IPR_2}^2}$, normalized by its maximal value, giving the same critical values $W_c=0.5$ and $\nu \approx 1$ as in Fig.~\ref{fig:QREM_EE}.}
\end{figure}

For the latter one, one can observe the parametric difference between these cases as the leading in $L$ solution is
\be
\lrp{\sigma W}^c \simeq e L \ln (e L) \lra W = \frac{2}{\sqrt{\pi}} L^{1/c-1/2}\lrb{e \ln (eL)}^{1/c} \ .
\ee
This gives the standard result for the Hermitian case
\be
W_H \sim L^{1/2} \ln L
\ee
and the completely unexpected one for the gain-loss non-Hermitian one:
\be
W_{nH} \sim \lrp{\ln L}^{1/2} \ .
\ee

To sum up, we have demonstrated that in the limit of large $L$, $W_{nH}\ll W_H$ and their ratio goes to zero.

The latter consideration is consistent with the generic consideration of the Hermitian case, where the ratio $W_0/t$ of the standard deviation of the diagonal $\mean{E_n^2}^{1/2} = W_0$ and the hopping $t$ terms at the critical point is given by the function of the vertex degree $d$
\be
\lrp{\frac{W_0}{t}} = d \ln d \ .
\ee
This is related to the first at large $W_0$, therefore the probability to have a resonance energy is given by the first power of $t/W_0$.
In the non-Hermitian case with the same variances for real and imaginary parts, $\mean{E_{n,R}^2}^{1/2} = \mean{E_{n,I}^2}^{1/2} = W_0$, the above probability of having small $|E_n-E_0|$ is given by $(t/W_0)^2$ as each of real and imaginary parts should be in the vicinity $t$ from $E_{0,R/I}$, respectively.
Thus, in this case
\be
\lrp{\frac{W_0}{t}}^2 = d \ln d \ .
\ee

This is consistent with both the above results, where $d = L$, $t=1$, and $W_0 = \sigma W = \sqrt{\pi L}W/2$, and the ones of the Rosenzweig-Porter model~\cite{DeTomasi2022nonHerm_RP}, where $d=N$, $t=N^{-\gamma/2}$, $W_0 = 1$.

%%%%%%%%%%%%%%%%%%%%%%%%%%%%%
\subsection{Numerical Result for the non-Hermitian QREM}\label{Subsec:QREM_numerics}
%%%%%%%%%%%%%%%%%%%%%%%%%%%%%
After providing an analytical perspective on the non-Hermitian QREM, which reveals that the model is substantially more localized compared to its Hermitian counterpart, we proceed to the numerical validation of our hypotheses. The probes employed for this validation are the same as those utilized for the random-field Heisenberg model, %as detailed in
Eq.~\eqref{eq:H_moni}, that serve to distinguish extended and localized phases.

Figures~\ref{fig:QREM_EE} and~\ref{fig:QREM_I} focus on the entanglement entropy and fractal dimension, respectively, and their associated fluctuations. Within both figures, panels (a) and (c) display the raw data as a function of $W$ across various system sizes. As expected, the range in which the model exhibits delocalization denoted  $S/S_{Page}\rightarrow 1$ and $D_2\rightarrow 1$, broadens with the system size $L$. Fingerprints of the transition are given to the moving maximum of the fluctuations of $S$ and $D_2$ visible in Fig.~\ref{fig:QREM_EE}(c) and Fig.~\ref{fig:QREM_I}(c), respectively. To find the scaling of the value of the critical point with system size, we perform a finite-size scaling analysis, which is shown in Figs~\ref{fig:QREM_EE}(b),~(d) and Figs.~\ref{fig:QREM_I}(b),~(d). As expected from our analytical considerations, we found with $W_c\sim \ln^{1/2}{L}$ a good collapse of the curves.

%Conversely, panels (b) and (d) in Figs~\ref{fig:QREM_EE} and ~\ref{fig:QREM_I} offer a finite-size consolidation of the curves, informed by our analytical insights suggesting that $W_c \sim \ln{L}$.

%%%%%%%%%%%%%%%%%%%%%%%%%%%%%
\section{Outlook $\&$ Conclusion}\label{Sec:concl}
%%%%%%%%%%%%%%%%%%%%%%%%%%%%%
In this work, we examine the phase of the random-field Heisenberg model under monitoring in the no-click limit. Without monitoring, the model is believed to exhibit a many-body localization transition, which separates an ergodic phase at weak disorder from a localized one at strong disorder.

Under random space continuous monitoring in the no-click limit, the model is mapped to a non-Hermitian system with complex random fields. Numerically, we investigate the localization properties of the model, in particular, the entanglement entropy and the fractal dimension in the Fock space. We found that the model exhibits a more robust form of localization. We provide a finite-size scale analysis, which suggests that the extended phase disappears in the thermodynamic limit, leaving the system in MBL for any finite amount of disorder. It is important to note that in the case of continuous but non-random measurements in space, the model is expected to exhibit the typical MBL  transition, as demonstrated in Ref.~\onlinecite{Kazuki_2023}.

Although we do not have an analytical approach for our random field Heidelberg model, we consider the non-Hermitian analog of QREM as a toy and analytically tractable model to test our conjectures. In the past, the Hermitian QREM has been used to describe the MBL transition, providing an evidence of MBL and the existence of many-body mobility edges. We consider a non-Hermitian version of the QREM, which mimics our random gain and loss terms. Using a self-consistent theory of localization, we show that the non-Hermitian QREM exhibits sharp many-body mobility edges, and the system is parametrically more localized than its corresponding Hermitian counterpart.

Our work establishes the foundation for stabilizing phases of matter through the application of random gain and loss terms. Future studies will investigate the potential of this measurement protocol to stabilize many-body localized topological phases and to induce localization in systems with higher dimensions and long-range interactions. Additionally, promising avenues of research include examining the effects of non-Hermitian elements on the avalanche theory of delocalization, as well as their influence on real-time dynamics~\cite{Turkeshi_2021,Turkeshi_2023}.

\begin{acknowledgments}
We are grateful to J.~Bardarson, D.~Belkin, B.~Clark, R.~Hamazaki, M.~Ippoliti, M.~Schiro, K.~Su, and S.~Vijay, for stimulating discussion. We also express our gratitude to
R. Hamazaki  for comments and a  critical reading of the manuscript.
I.~M.~K. acknowledges the support
%by Russian Science Foundation (Grant No. 21-12-00409).
by the European Research Council under the European
Union's Seventh Framework Program Synergy ERC-2018-SyG HERO-810451.
G.~D.~T. acknowledges the support from the EPiQS Program of the Gordon and Betty Moore Foundation.
\end{acknowledgments}

\bibliography{Lib}

\end{document}